\documentstyle{elsart}     
\begin{document}

\renewcommand{\theequation}{\thesection.\arabic{equation}}
\def\ade{$A$--$D$--$E$\space}
\def\weight#1#2#3#4#5{#1\!\!\left(\matrix{#5&#4\cr #2&#3\cr}\right)}
\def\wt#1#2#3#4#5#6{#1\!\!\mbox{
$\left(\matrix{#5&#4\cr#2&#3\cr}\biggm|\mbox{$#6$}\right)$}}
\def\cell#1#2#3#4#5#6{C\!\!\left(\matrix{#6&#5&#4\cr #1&#2&#3\cr}\right)}
\def\Wf#1#2#3#4#5#6#7#8#9{W_{p\times q}\mbox{$\left(
   \matrix{#5&#8&#4\cr#9&&#7\cr #1&#6&#3\cr}\biggm|\mbox{$#2$}\right)$}}
\def\Wfml#1#2#3#4#5#6#7#8#9{W_{p\times l}\mbox{$\left(
   \matrix{#5&#8&#4\cr#9&&#7\cr #1&#6&#3\cr}\biggm|\mbox{$#2$}\right)$}}
\def\Wfln#1#2#3#4#5#6#7#8#9{W_{l\times q}\mbox{$\left(
   \matrix{#5&#8&#4\cr#9&&#7\cr #1&#6&#3\cr}\biggm|\mbox{$#2$}\right)$}}

\newcommand{\hs}[1]{\mbox{\hspace*{#1cm}}}

\def\ZZ{{\Z}}
\def\and{\;\;{\rm and}\;\;}
\def\for{\;\;{\rm for}\;\;}
\def\re{\mbox{${\Re}e$ }}
\def\im{\mbox{${\Im}m$ }}
\def\i{\mbox{\small i}}
\def\-{\!-\!}
\def\+{\!+\!}
\def\<{\langle}
\def\>{\rangle}
\def\({\biggl(}
\def\){\biggr)}
\def\h{\hspace*{0.5cm}}
\def\mh{\hspace*{-0.5cm}}
\def\half {\mbox{$\textstyle {1 \over 2}$}}
\def\talf {\mbox{$\textstyle {3 \over 2}$}}
\def\mat {\pmatrix}
\def\smat#1{\mbox{\small $\mat{#1}$}}
\def\ba{\begin{array}}
\def\ea{\end{array}}
\def\be{\begin{eqnarray}}
\def\ee{\end{eqnarray}} 
\def\beq{\begin{equation}}
\def\eeq{\end{equation}} 
\def\bra#1{\langle #1|}
\def\ket#1{|#1\rangle}
\def\ol{\overline}
\def\no{\nonumber}
\def\disp{\displaystyle}
\def\aa{\overline{a}}
\def\bb{\overline{b}}
\def\cc{\overline{c}}
\def\reone{\mbox{regime {\sl III/IV} }}
\def\retwo{\mbox{regime {\sl I/II} }}
\def\T{\mbox{\boldmath {$T$}}}
\def\Y{\mbox{\boldmath { $ Y$}}}
\def\t{\mbox{\boldmath {$ t$}}}
\def\1{\mbox{\boldmath I}}
\def\C{\mbox{\boldmath {$\cal C$}}}
\def\B{\mbox{\boldmath {$B$}}}
\def\a{\mbox{$\aphaa$}}
\def\U{\mbox{$\cal A$}}
\def\e{\mbox{$\epsilon$}}
\def\b{\mbox{$\beta$}}
\def\la{\mbox{${\ell}\!a$}}
\def\Le{\mbox{${\ell}\!e$}}
\def\lA{\mbox{${\ell}\!A$}}
\def\[{\mbox{\huge [}}
\def\]{\mbox{\huge ]}}
\def\lam{\lambda}

\begin{frontmatter}
\title{Critical behaviour of the dilute O($n$), Izergin-Korepin\\
and dilute $A_L$ face models: Bulk properties}

\author{Y.K. Zhou} and 
\author{M.T. Batchelor}
\address{Department of Mathematics, School of Mathematical
Sciences,\\ The Australian National University, Canberra ACT 0200,
Australia}

\begin{abstract}
The analytic, nonlinear integral equation approach is 
used to calculate the finite-size corrections to the transfer 
matrix eigen-spectra of the critical dilute O($n$) model on 
the square periodic lattice. The resulting bulk conformal 
weights extend previous exact results obtained in the 
honeycomb limit and include the negative spectral parameter regimes. The
results give the operator content of the 19-vertex Izergin-Korepin model
along with the conformal weights of the dilute
$A_L$ face models in all four regimes.

\end{abstract}

\end{frontmatter}

\section{Introduction}
\setcounter{equation}{0}

Among other physical phenomena, 
the integrable dilute O($n$) model on the square lattice 
\cite{N:90a} is relevant to self-avoiding polymer chains in the 
bulk \cite{N:90b,N:90c}.
The partition function of the dilute O($n$) model is defined 
by \cite{N:90a,WN:93}  
\beq
Z = \sum_{{\mathcal G}} \rho_1^{m_1} \cdots \rho_9^{m_9} \; n^{P},
\label{Z}
\eeq
where the sum is over all configurations 
${\mathcal G}$ of non-intersecting closed
loops which cover some (or none) of the lattice bonds.
The possible loop configurations at a vertex are 
shown in Fig. 1, with a vertex of type
$i$ carrying a Boltzmann weight $\rho_i$. 
In configuration ${\mathcal G}$,
$m_i$ is the number of occurrences of the 
vertex of type $i$, while $P$ is the
total number of closed loops of fugacity $n$.

The loop weights in (\ref{Z}) are \cite{N:90a,WN:93} 
\be
\rho_1 (u)&=&1 + {\sin u \sin (3\lam-u)
                        \over \sin 2\lam\sin 3\lam}  \no\\
\rho_2 (u)&=&\rho_3 (u) = {\sin (3\lam-u)\over \sin 3\lam}\no \\
\rho_4 (u)&=&\rho_5 (u) = {\sin u \over \sin 3\lam} \no \\
\rho_6 (u)&=&\rho_7 (u) = {\sin u \sin (3\lam-u) \over \sin
                                     2\lam\sin 3\lam}  \\
\rho_8 (u)&=&{\sin (2\lam-u) \sin (3\lam-u)\over\sin
                                     2\lam\sin 3\lam} \no \\
\rho_9 (u)&=&-{\sin u \sin(\lam-u) \over \sin 2\lam\sin 3\lam}. \no
\ee
Here $n=-2\cos 4\lam$.
These weights were originally constructed via a mapping involving the
Potts model \cite{N:90a} and later seen to satisfy the 
Yang-Baxter equation for loop 
models \cite{WN:93,NWB:93}. 
On the other hand, when mapped to a 3-state vertex model \cite{N:90a}, 
the dilute O($n$) model is seen to be related to the integrable  
19-vertex model of Izergin and Korepin \cite{IK:81}. 
The Nienhuis O($n$) model on the honeycomb lattice 
\cite{N:82,N:84,Bax:86} 
follows from either of the special values
$u=\lam$ and $u=2\lam$ of the spectral parameter \cite{N:90a,R:91}. 
In the appropriate region the model thus contains the essential physics
of the self-avoiding polymer problem at 
$n=0$ \cite{N:90b,N:82,N:84,N:87,D:90a}. 

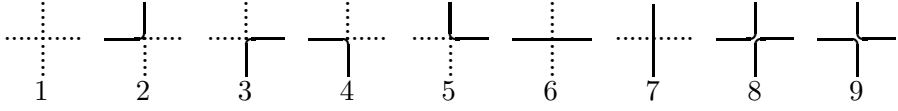
\begin{figure}[t]
\begin{center}
\setlength{\unitlength}{0.01250000in}
\begin{picture}(34,34)(88,673)
\multiput( 90,690)(3,0){11}{.}
\multiput(105,705)(0,-3){11}{.}
\put(103,665){\small$1$}
\end{picture}
\setlength{\unitlength}{0.01250000in}
\begin{picture}(34,34)(88,673)
\put(106.5,706){\thicklines\line(0,-1){13}}
\put(90,690.5){\thicklines\line(1,0){12.5}}
\put(102.5,694.6){\thicklines\oval(8,8)[br]}
\multiput(119.5,690)(-3,0){6}{.}
\multiput(105,675)(0,3){6}{.}
\put(103,665){\small$2$}\end{picture}
\setlength{\unitlength}{0.01250000in}
\begin{picture}(34,34)(88,673)
\put(110.5,686.7){\thicklines\oval(8,8)[tl]}
\put(122.5,690.8){\thicklines\line(-1,0){12}}
\put(106.5,675){\thicklines\line(0,1){12.5}}
\multiput(90,690)(3,0){6}{.}
\multiput(104.6,705)(0,-3){6}{.}
\put(103,665){\small$3$}\end{picture}
\setlength{\unitlength}{0.01250000in}
\begin{picture}(34,34)(88,673)
\put(106.5,675){\thicklines\line(0,1){13}}
\put(90,690.5){\thicklines\line(1,0){12.5}}
\put(102.5,686.5){\thicklines\oval(8,8)[tr]}
\multiput(119.5,690)(-3,0){6}{.}
\multiput(104.6,705)(0,-3){6}{.}
\put(103,665){\small$4$}\end{picture}
\setlength{\unitlength}{0.01250000in}
\begin{picture}(34,34)(88,673)
\put(110.5,694.7){\thicklines\oval(8,8)[bl]}
\put(122.5,690.8){\thicklines\line(-1,0){12}}
\put(106.5,706){\thicklines\line(0,-1){12.5}}
\multiput(90,690)(3,0){6}{.}
\multiput(105,675)(0,3){6}{.}
\put(103,665){\small$5$}\end{picture}
\setlength{\unitlength}{0.01250000in}
\begin{picture}(34,34)(88,673)
\put( 90,690.5){\thicklines\line(1,0){33}}
\multiput(105,705)(0,-3){11}{.}
\put(103,665){\small$6$}
\end{picture}
\setlength{\unitlength}{0.01250000in}
\begin{picture}(34,34)(88,673)
\put(106.6,705){\thicklines\line(0,-1){30}}
\multiput( 90,690)(3,0){11}{.}
\put(103,665){\small$7$}
\end{picture}
\setlength{\unitlength}{0.01250000in}
\begin{picture}(34,34)(88,673)
\put(106.5,706){\thicklines\line(0,-1){13}}
\put(90,690.5){\thicklines\line(1,0){12.5}}
\put(102.5,694.6){\thicklines\oval(8,8)[br]}
\put(110.5,686.7){\thicklines\oval(8,8)[tl]}
\put(122.5,690.8){\thicklines\line(-1,0){12}}
\put(106.5,675){\thicklines\line(0,1){12.5}}
\put(103,665){\small$8$}
\end{picture}
\setlength{\unitlength}{0.01250000in}
\begin{picture}(34,34)(88,673)
\put(106.5,675){\thicklines\line(0,1){13}}
\put(90,690.5){\thicklines\line(1,0){12.5}}
\put(102.5,686.5){\thicklines\oval(8,8)[tr]}
\put(110.5,694.7){\thicklines\oval(8,8)[bl]}
\put(122.5,690.8){\thicklines\line(-1,0){12}}
\put(106.5,706){\thicklines\line(0,-1){12.5}}
\put(103,665){\small$9$}
\end{picture}
\vskip 5mm
\caption{The 9 vertices of the dilute O($n$) model.}
\vskip 5mm
\end{center}
\end{figure}

The dilute O($n$) model has recently been used to 
construct a family of dilute \ade lattice models 
\cite{WNS:92,Roche:92,WNS:93}. These models
are restricted solid--on--solid
models with a finite number of heights 
built on the \ade Dynkin diagram. At criticality, 
the face weights are \cite{WNS:92,Roche:92,WNS:93}
\be
& & \wt Wabcdu
= \;\rho_1(u)
  \delta_{a,b,c,d} +\rho_2(u) \delta_{a,b,c} A_{a,d}+\rho_3(u)
\delta_{a,c,d} A_{a,b} \no\\*
& &\quad \mbox{} +\sqrt{S_a \over S_b}\rho_4 (u) \delta_{b,c,d} A_{a,b}
     +\sqrt{S_c \over S_a}\rho_5(u) \delta_{a,b,d} A_{a,c}
     +\rho_6(u) \delta_{a,b} \delta_{c,d} A_{a,c}  \label{adeface-d}\\
& &\quad \mbox{}  +\rho_7(u) \delta_{a,d} \delta_{c,b} A_{a,b} +\rho_8(u)
\delta_{a,c} A_{a,b} A_{a,d} +
       \sqrt{{S_a S_c \over S_b S_d}} \rho_9(u) \delta_{b,d} A_{a,b}
A_{b,c} \no
\ee
where the $\rho_i$ are as given above.  
The generalized Kronecker
delta is unity if all its arguments take the same
value and is zero otherwise. 
The Perron-Frobenius vectors $S_{a}$ in the face weights are
the eigenvector of the largest eigenvalue of the
adjacency matrix $A$ of the \ade graphs,
\beq
\sum_{b}A_{a,b}S_b=2\cos{\pi\over L+1} S_a\;,
\eeq
where for the dilute $A_L$ models, $L$
is the number of graph states, with $a,b,c,d=1,2,\cdots,L$.

The dilute O($n$) model exhibits various branches of critical 
behaviour \cite{BN:89,BNW:89,WBN:92,Warnaar}. 
These are reflected in the properties
of the dilute \ade models, for which there are four physical
branches \cite{WNS:92}  
\be
&\mbox{branch {\it 1}}
\hs{1.8} 0<u< 3\lam &\hs{0.3}\lam={\pi\over 4}{L\over L+1} \hs{0.7}
                L=2,3,\cdots\no\\
&\mbox{branch {\it 2}}
\hs{1.8} 0<u< 3\lam &\hs{0.3}\lam={\pi\over 4}{L+2\over L+1}\hs{0.7}
                  L=3,4,\cdots\no\\
&\mbox{branch {\it 3}}
\h -\pi+3\lam<u<0 &\hs{0.3}\lam={\pi\over 4}{L+2\over L+1}\hs{0.7}
                   L=3,4,\cdots \label{branches}\\
&\mbox{branch {\it 4}}
\h -\pi+3\lam<u<0  &\hs{0.3}\lam={\pi\over 4}{L\over L+1}\hs{0.7}
                  L=2,3,\cdots\no
\ee

Recent studies have highlighted the prominence of the dilute $A_L$ 
face models, which admit an off-critical extension \cite{WNS:92,WNS:93}. 
In regime 2
the $A_3$ model lies in the same universality class as the Ising model
in a magnetic field and gives the magnetic exponent 
$\delta=15$ \cite{WNS:92,WNS:93,WPSN:94}. This $A_3$ model also shows
the $E_8$ scattering theory for massive excitations over the 
groundstate \cite{BNW:94,WP:94,GN:95}. Both $su(2)$ and $su(3)$ fusion 
hierarchies of the dilute $A_L$ face models have
been constructed in \cite{Zf:96,ZPG:95}.

In this paper we both generalise and extend earlier calculations
of the critical properties, such as the central charges and bulk
scaling dimensions (the conformal spectra), of the 
dilute O($n$) model and the related dilute $A_L$ and Izergin-Korepin models. 
After outlining the necessary
preliminaries in Section 2, our calculations are presented in
Section 3 for branches 1 and 2 and in Section 4 for branches 3 and 4.
The method employed involves the extension of the nonlinear integral
equation approach \cite{KB:90,KBP:91,WBN:92} 
to obtain the complete conformal spectra, as has been done for
the six-vertex model \cite{KWZ:93,F:95} 
and most recently \cite{Zhou:95} for the 
Andrews-Baxter-Forrester (ABF) model \cite{ABF:84}.
Having read Section 2, those readers not specifically interested in the
technical details may prefer to skip to Section 5 where a   
discussion of our results for the various models concludes the paper.

\section{Bethe equations and known results}
\setcounter{equation}{0}

As we are interested in bulk critical behaviour, we 
consider periodic boundary conditions across a finite lattice of width $N$, 
where for convenience we take $N$ even.
The eigenvalues $T(u)$ for the row-transfer matrix $\T(u)$ of the 
dilute O($n$) model are given by \cite{BNW:89,WBN:92,Warnaar}
\be
T(u) & = & e^{-\i \phi}\,\frac{s(2\lam-u)s(3\lam-u)}{s(2\lam)s(3\lam)}\,
           \frac{Q(u+\lam)}{Q(u-\lam)}  \no \\
     && \qquad +     \frac{s(u)s(3\lam -u)}{s(2\lam)s(3\lam)}\,
           \frac{Q(u)Q(u-3\lam)}{Q(u-\lam)Q(u-2\lam)} \no\\*
     && \qquad\qquad + e^{\i \phi}\,
           \frac{s(u)s(\lam-u)}{s(2\lam)s(3\lam)}\,
           \frac{Q(u-4\lam)}{Q(u-2\lam)}, \label{BAE}
\ee
where
\beq
s(u)=\sin^{N}(u), \qquad
Q(u)=\prod_{j=1}^{m} \cosh(\i u-u_j)
\eeq
and the $m$ zeros $\{u_j\}$ satisfy the Bethe equations
\beq
e^{\i \phi} \left[\frac{\cosh(u_j+\i\lam)}{\cosh(u_j-\i\lam)}\right]^N  = 
- \prod_{k=1}^{m} \frac{\sinh(u_j-u_k+2\i\lam)\sinh(u_j-u_k-\i\lam)}
                       {\sinh(u_j-u_k-2\i\lam)\sinh(u_j-u_k+\i\lam)} 
\label{bethe}
\eeq
for $j=1,\ldots,m$. It is convenient to label the sectors of $\T(u)$ by 
$\ell = N - m$, where $\ell=0$ for the largest (groundstate) sector, 
$\ell=1$ for the next largest, etc.

The Bethe equations ensure that the eigenvalues $T(u)$ are analytic
functions of $u$.
Apart from the phase factors $\phi$ these equations are
the Bethe equations of the Izergin-Korepin 
model \cite{R:83,VR:83,T:88}.
In general $\phi$ is a continuous variable associated 
with a ``seam'' to ensure that loops which wrap round the 
cylinder carry the correct weight $n$. Thus
\beq
\phi = \pi - 4\lam \label{on} 
\eeq
for the dilute O($n$) model in the largest ($\ell=0$) sector of $\T(u)$ 
with $\phi=0$ in all other sectors. For the Izergin-Korepin model, 
$\phi=0$ in all sectors. 

On the other hand, for the dilute $A_L$ face models there is a fixed number 
of Bethe roots ($\ell=0$) and  
\beq
\phi = \pi s/(L+1) \label{ph} 
\eeq
with $s=1,\ldots,L$.
In this case the transfer matrix $\T(u)$ has elements 
\beq
\langle \sigma | \T(u) | \sigma' \rangle =
\prod_{j=1}^{N} \wt W{\sigma_j}{\sigma_{j+1}}{\sigma_{j+1}'}{\sigma_j'}u,
\eeq
where the paths $\sigma=\{\sigma_1,\sigma_2,\ldots,\sigma_N\}$ and
$\sigma'=\{\sigma_1',\sigma_2',\ldots,\sigma_N'\}$
are allowed configurations of heights along a row with periodic
boundary conditions $\sigma_{N+1}=\sigma_1$ and $\sigma_{N+1}'=\sigma_1'$.
The face weights are those defined in (\ref{adeface-d}).
The expression for the eigenvalues $T(u)$ was given 
in \cite{BNW:94} in terms of the more general elliptic functions.
As we are also interested in the dilute O($n$) model, we
do not restrict the crossing parameter $\lam$ to the values 
given in (\ref{branches}). This may lead to unphysical regimes in the 
dilute $A_L$ face model for which, however, the finite-size
corrections to the transfer matrix eigenspectra are still of interest
from the viewpoint of 
statistical mechanics and conformal field theory.

\subsection{Central charge}

Some exact results are known for the dilute O($n$) 
model \cite{BNW:89,WBN:92,Warnaar}. In particular, the 
central charge is found to be 
\be
c &=& 1 - \frac{3 \phi^2}{\pi (\pi - 2 \lambda)} 
\qquad  \mbox{branches 1 \& 2}, \label{c1&2}\\ 
c &=& \frac{3}{2} - \frac{3 \phi^2}{2 \pi \lambda} 
\qquad\qquad\,\; \mbox{branches 3 \& 4}.
\label{c3&4}
\ee
These results follow from the finite-size behaviour
of the largest eigenvalue. The result (\ref{c1&2}) had already been obtained
from the Bethe equations in the honeycomb limit \cite{BB:88,Su:88,BB:89}.
However, the result (\ref{c3&4}) \cite{WBN:92} had to await the 
development of the more sophisticated nonlinear integral equation approach 
\cite{KB:90,KBP:91,WBN:92} (see also \cite{PK:91,KP:91}). 

The reason for this is that the distribution of Bethe roots for the largest
eigenvalue differs significantly
in each case. In the limit of infinite size $N$        
the Bethe roots are distributed on the lines \cite{BNW:89,WBN:92,Warnaar}
\be
&\mbox{branches 1 and 2}& \quad \im (u_j)=\half\pi, \label{good} \\
&\mbox{branches 3 and 4}& \quad \im (u_j)=\pm\half\pi \lam. \label{bad}
\ee
Whereas there are no finite-size deviations from the line (\ref{good}),
the finite-size deviations from (\ref{bad}) are severe
enough to render the more standard root density approach\footnote{See, for
example, \cite{SNW:92} and references therein.} invalid.  
Here we extend the analytic, nonlinear integral equation 
approach in the dilute O($n$) model 
\cite{WBN:92,Warnaar} to the calculation of the 
conformal weights in all four branches. Our treatment follows that given 
in the recent study of the ABF model \cite{Zhou:95}. 

The above results for the central charge have
already been used to obtain the central charges of the dilute \ade 
models \cite{WNS:92,WNS:93,Warnaar}.    
In particular, for the dilute $A_L$ face models, either (\ref{ph}) or 
the O($n$) value (\ref{on}) with
the appropriate value of $\lam$ in (\ref{branches}) gives
\be
c=\cases{1-\disp{6\over h(h-1)},    &branches 1 \& 2, \cr
        \talf-\disp{6\over h(h-1)}, &branches 3 \& 4, }\label{c}
\ee
where
\beq
h=\cases{L+1, &branches 2 \& 4, \cr
         L+2, &branches 1 \& 3. }
\eeq
The first two branches give realisations of the unitary 
minimal series, while the other two branches involve a product of the 
unitary minimal series and an Ising model. 

The O($n$) model had earlier
been identified \cite{DF:84,SS:85} in the conformal classification 
scheme \cite{BPZ:84,FQS:84} 
with the aid of the Nienhuis Coulomb gas results \cite{N:82,N:84}. 
In particular, 
$
c = 1 - 6(g-1)^2/g,
$
where $g \in [1,2]$, with $g=h/(h-1)$, in the high temperature phase 
(branch 1) and $g \in [0,1]$, with $g=(h-1)/h$, in the low temperature 
phase (branch 2). Here $g = 2 (1-2 \lambda/\pi)$. 
The Ising value $c=\half$ thus occurs both for the 
dilute $A_2$ model ($n=1$ in the high temperature O($n$) 
phase) and the dilute $A_3$ model ($n=\sqrt 2$ in the low temperature O($n$) 
phase). The central charges of the dilute $A_L$ face models have recently 
been estimated numerically from the 
finite-size diagonalisation of the dilute $A_L$ model transfer matrix for 
various $L$  on all four branches \cite{OP:95}. 
The central charge has also been derived by solving the
transfer matrix functional relations of the dilute $A_L$ model on branches  
2 and 4 \cite{ZP:95p}. 
The calculation confirms the result (\ref{c}) obtained 
via the dilute O($n$) model \cite{WNS:92,WNS:93,Warnaar}.

\subsection{Scaling dimensions}

Various scaling dimensions have been calculated 
via the Bethe equations for the 
dilute O($n$) model.
Again in the honeycomb limit for branches 1 and 2, 
the `magnetic' set of scaling 
dimensions is found to be \cite{BB:88,Su:88,BB:89}
\beq
X_\ell^\sigma = \frac{\ell^2 (\pi-2\lam)^2 -(\pi-4\lam)^2}{4\pi(\pi-2\lam)} 
   = {\mbox{$\textstyle {1 \over 8}$}} g\, \ell^2 - \frac{(g-1)^2}{2 g}.
\label{mag12}
\eeq
Alternatively, this result is written as 
\beq
X_\ell^\sigma =\cases{2 \Delta_{\ell/2,0}\,, &branch 1 \cr
                      2 \Delta_{0,\ell/2}\,, &branch 2 }
\eeq
where
\beq
\Delta_{r,s}^{(h)}={[h r-(h-1)s]^2-1\over 4h(h-1)}
\label{Delta}
\eeq
is the Kac formula. This result had earlier been obtained via 
Coulomb gas calculations \cite{Sal:86,D:87}.

On the other hand, both numerical evidence \cite{BB:89} and 
root-density calculations \cite{SNW:92} revealed the set of `thermal'
dimensions to be 
\beq
X_j^\epsilon = \frac{j^2 \pi - j (\pi-4\lam)}{\pi-2\lam} 
             = \frac{2}{g} j (j+1) - 2 j.
\label{therm12}
\eeq 
Both the results (\ref{mag12}) and (\ref{therm12}) 
generalized earlier results via the  
Coulomb gas \cite{N:82,N:84,N:87}.\footnote{The leading thermal
dimension had been conjectured earlier for the O($n$) model by 
Cardy and Hamber \cite{CH:80}.} 
The thermal dimensions follow from 
\beq
X_j^\epsilon =\cases{2 \Delta_{1,2j+1}\,, &branch 1 \cr
                      2 \Delta_{2j+1,1}\,, &branch 2 }
\eeq
in the Kac formula \cite{BB:89}.

On the other hand, the situation is not so clear for branches 3 and 4 
of the dilute square lattice model. Numerical evidence \cite{WBN:92}
indicates that the magnetic dimensions are given by
\beq
X_\ell^{\sigma} = \frac{\lambda \ell^2}{2 \pi} - \frac{(\pi-4\lambda)^2}
{8 \pi \lambda}. 
\label{mag34}
\eeq
The only known thermal result is $X_1^\epsilon = 1$ \cite{WBN:92}. 
There are no Coulomb gas results for these branches.

The conformal weights of the dilute $A_L$ face models have been
estimated numerically from the finite-size diagonalisation
of the transfer matrix (for $L=3$ and $L=4$ at $u=3\lambda/2$ on 
all four branches) \cite{OP:95} and from numerical 
solutions to the Bethe equations for $L=3$ \cite{GN:95}. 
For branches 1 and 2,
the results fulfil the expectation that the scaling dimensions 
reflect the conformal weights of the unitary minimal series. For 
branches 3 and 4, they reflect a product of the Ising and unitary minimal 
series. The related modular invariant partition function has been
discussed at length in \cite{OP:95}. 

As mentioned above, the finite-size corrections 
to the transfer matrix eigenspectra have been obtained for 
the dilute $A_L$ face models in branches 2 and 4 in \cite{ZP:95p}
via the functional relation method \cite{PK:91,KP:91,KlPe:92}. 
The analytic calculation confirms the conformal weights obtained via
the calculation of the local height 
probabilities for $L$ odd \cite{WPSN:94}. Here we  
consider the dilute models in all four branches with more general crossing
parameter $\lam$ and calculate the conformal
spectra for each branch.

\section{Branches 1 and 2}\setcounter{equation}{0}

We consider branches 1 and 2 defined by 
\beq
0<u<3\lam \h\h {\pi/6}\le\lam<{\pi/ 3}.
\eeq
This regime covers the $\lam$ values (\ref{branches}) for 
the dilute $A_L$ models.
However, the derivation below is also valid for the dilute O($n$) 
and Izergin-Korepin models in the 
larger interval $0<\lam<{\pi/3}$.
Let us introduce the new variable $v=\i u$ 
with a shift $v_j=u_j-\half\i\pi$. The Bethe  equations
(\ref{bethe}) are then of the form
\beq
p(v_j)=-1,
\eeq
where
\be
p(v)&=&e^{-\i\phi} {\Phi(v- \i \lam)q(v-\i\lam)q(v+2\i\lam)\over
  \Phi(v+\i \lam)q(v+\i\lam)q(v-2\i\lam)}, \\
\Phi(v)&=&\sinh^N v\;,  \h\h  q(v)
 \;=\;\prod_{j=1}^{m}\sinh(v-v_j)\; . \label{phiq}
\ee
After the shift, the Bethe roots $v_j$ are 
distributed  along the real axis, with 
\beq
\ol{q}(v)=q(\ol{v}) \and \ol{p}(v)=1/p({v}) \label{qp}
\eeq
where the overbar denotes complex conjugation.

\subsection{Nonlinear integral equation}

Define two functions that
are Analytic and Non-Zero ({\sc anz}) in the strips around the real axis:
\be
\alpha(v)&=&e^{\i\omega}
\;g(v)\;p(v+\i\xi),  \no \\
A(v)&=& 1+\alpha(v)/g(v).
\ee
The phase factor $\omega$ has been introduced for 
taking different branches of the log function involved in the
subsequent Fourier transforms. We take 
\beq
\omega=\cases{
 {\rm sgn}(v)(\ell-r)\pi &dilute O($n$) model, $\ell\ne 0$ \cr
 r\pi &dilute O($n$) model, $\ell= 0$ \cr
 \pi(r-s)  & dilute $A_L$ face model \cr}
\label{cases}
\eeq
where the function ${\rm sgn}(v)=-1$ for $Re(v)>0$ 
and $+1$ otherwise. For the O$(n)$ model
the integers $r,s$ are restricted (as discussed further in
section~5). For the moment we leave them as arbitrary integers.
The function $g$ is introduced for compensating
the anticipated bulk behaviour of $p(v+\i\xi)$ and is given by
\beq
g(v)=\left({{\rm th}\rho(v+\i\lam+\i\xi)\over{\rm th}\rho(v-\i\lam+\i\xi)}
     \right) ,\h\h \rho=\pi/6\lam,
\eeq
where $0<\xi\le\half\pi$. The $\i\pi-$periodic function $\alpha$ 
can be rewritten as 
\beq
\alpha(v)= e^{\i\omega-\i\phi} g(v){\Phi(v- \i \lam+\i\xi)
  q(v-\i\lam+\i\pi+\i\xi)q(v+2\i\lam+\i\xi)\over
  \Phi(v+\i \lam+\i\xi)q(v+\i\lam+\i\xi)q(v-2\i\lam+\i\pi+\i\xi)}.
\label{al}\eeq
The above treatment results in the function $\alpha$ representing 
finite-size corrections. To see this we 
consider the Fourier transform pair
\be
\alpha(k)&=&{1\over 2\pi}\int_{-\infty}^{\infty}
          \(\ln\alpha(v)\)^{\prime\prime}e^{-\i kv}\;dv \no \\
\(\ln\alpha(v)\)^{\prime\prime}
 &=&\int_{-\infty}^{\infty}\alpha(k)e^{\i kv}\;dk
\label{Fou}
\ee
for $\alpha$ and similarly for $A$. The Fourier transform of $q(v)$ is 
defined to be
\be
&q(k)={1\over 2\pi}\disp{\int_{-\infty+{\rm i} r}^{\infty+\i r}}
          \(\ln q(v)\)^{\prime\prime}e^{-{\rm i} kv}\;dv \h &0<r<\pi,\no \\
&\(\ln q(v)\)^{\prime\prime}=\disp{\int_{-\infty}^{\infty}}
   q(k)e^{\i kv}\;dk\h\h &0<\im(v)<\pi.
\ee
To represent $\alpha(k)$ by $A(k)$ and $\ol{A}(k)$ we also need 
another relation, which can be given by applying Cauchy's theorem 
to the auxiliary function
\beq
h(v)={1+p(v)\over p(v)q(v)}, \label{h} 
\eeq
which satisfies the non-trivial analyticity property
\beq
\disp{\int_{-\infty+{\rm i}\xi}^{\infty+{\rm i}\xi}}
\(\ln h(v)\)^{\prime\prime}e^{-{\rm i} kv}\;dv=
  \disp{\int_{-\infty-{\rm i}\xi}^{\infty-{\rm i}\xi}}
 \(\ln h(v)\)^{\prime\prime}e^{-{\rm i} kv}\;dv\;.
\label{h1}\eeq
From the equations following Fourier transforming (\ref{al})
and inserting (\ref{h}) into (\ref{h1}) we obtain
\be
q(k)&=&{Nke^{(\half\pi k)}\cosh(\half\lam k)\over 
             2\sinh(\half\pi k)\cosh(\talf\lam k)} \no \\
 &&+{e^{(\half\pi k)}\cosh(\half\lam k)\over 
        2\sinh(\half\pi k-\lam k)\cosh(\talf\lam k)}
        \(e^{\xi k}A(k)-e^{-\xi k}\ol{A}(k)\),  \\
\alpha(k)&=&F(k)\;A(k)-F_\xi(k)\;\ol{A}(k),  
\ee
where
\beq
F_\xi(k)=-e^{-2\xi k}{\sinh(\lam k)\cosh(\half\pi k-\talf\lam k)\over
        \cosh(\talf\lam k)\sinh(\half\pi k-\lam k)} 
\eeq
and $F(k)=F_0(k)$.
Transforming back and integrating twice we obtain
the nonlinear integral equation
\beq
\ln\alpha(v)=F*\ln A-F_\xi *\ln \ol{A}+C +C' v,
\eeq
where the convolution  is defined by
\beq
(f*g)(v)=\int_{-\infty}^\infty f(w)g(v-w)\ dw.
\eeq
The constant $C'$ is chosen to be $C'=0$ for all terms to remain
finite. The other constant $C$ is scaling-dependent and is fixed
after taking the scaling limit defined by
\be
a_\pm(x)&=& \lim_{N\to\infty}\alpha(\pm v)/g(\pm v), \no \\
A_\pm(x)&=& \lim_{N\to\infty}A(\pm v)\;=\;1+a_\pm(x)\;.
\label{scaling12}
\ee
The nonlinear integral equation then becomes
\begin{equation}
\smat{\ln a_{\pm}\cr\ln \ol{a}_{\pm}}=2\i\sqrt{3}e^{-x}
 \smat{-e^{\mp 2\rho\i\xi}\cr e^{\pm 2\rho\i\xi}}
 +K*\smat{\ln A_\pm\cr\ln\ol{A}_\pm}+C_{\pm}\;\smat{1\cr -1}
  \; ,\label{a12}
\end{equation}
in which the kernel $K$ is given by
\beq
K=\smat{F_1&-F_2\cr -\ol{F}_2&\ol{F}_1},
\eeq
where
\begin{eqnarray}
F_1(x)&=&{F}_1(-x)={1\over 2\rho}F\(\pm{1\over 2\rho}x\),   \no \\
F_2(x)&=&\ol{F}_2(-x)={1\over 2\rho}F\(\pm{1\over 2\rho}x+2\i\xi\).
\end{eqnarray}  
We can see that $K^T(x)=K(-x)$, a key property to be used in
the derivation of the finite-size corrections. Taking $x\to\infty$ we obtain
\begin{equation}
C_\pm=\i\pi(\omega_\pm-\phi)/(\pi-2\lam),
\end{equation} 
where 
\beq
\omega_\pm=\cases{\omega\mp2\ell\lam   & dilute O($n$) model,\cr
\omega & dilute $A_L$ face model\cr}
\label{casespm}\eeq
The nonlinear integral equation (\ref{a12}) is equivalent to the Bethe  
equations for the largest eigenvalue, as given in
\cite{WBN:92,Warnaar}. The key difference is the change in the
integration constants $C_\pm$ for the low-lying excited states.
This is similar to the nonlinear integral equation 
approach in the six-vertex \cite{KWZ:93} and  ABF 
\cite{Zhou:95} models. In each case the constants contain 
the necessary information to extract the conformal weights.

\subsection{Conformal spectra}

The eigenvalues of the transfer matrix are given
by (\ref{BAE}). For small positive values of $u$ the first term in the
eigenvalue expression dominates exponentially. For small positive
imaginary $v$ we therefore have
\beq 
T(v)\sim e^{-\i\phi} \Phi(v-2\i\lam)
  \Phi(v-3\i\lam){q(v+\i\lam)\over q(v-\i\lam)}\;
\eeq
for the finite-size corrections. Taking Fourier transforms and integrating
twice yields
\be
\ln T(v)&=&-N f_\infty(v)+ {2\sqrt{3}\rho\over \pi}\int_{-\infty}^{\infty}
 \left({\sinh 4\rho(v-w-\i\xi)\over
  \sinh 6\rho(v-w-\i\xi)}\ln A(w)\right. \no \\
  &&\h\h \left.- {\sinh 4\rho(v-w+\i\xi)\over\sinh 6\rho(v-w+\i\xi)}\ln 
   \ol{A}(w)\right)dw, \label{t}
\ee
where the free energy is given by
\beq
 f_\infty(v)=2\int_{-\infty}^{\infty} dk\; {\sinh( k\i v)\sinh(3k\lam+\i vk)
   \cosh(5k\lam-k\pi)\cosh(k\lam)\over k\cosh(3k\lam)\sinh(k\pi)} . 
\eeq        
The integration constants have been fixed again by the limit $v\to\infty$.
Taking the thermodynamic limit $N\to\infty$ in (\ref{t})
and using the definitions
(\ref{scaling12}) gives
\be
\ln T(v)&=&-N f_\infty(v)+{2\sqrt{3}\i\rho\over N\pi}e^{2\rho v}
   \im\left(e^{-2\rho\i\xi} \int_{-\infty}^{\infty}
                        \ln A_+(x)e^{-x}\right) \no \\
  &&\h\h -{2\sqrt{3}\i\rho\over N\pi}e^{2\rho v}
 \im\left(e^{2\rho\i\xi} \int_{-\infty}^{\infty}\ln 
      {A}_-(x)e^{-x}\right)dx \label{T1}
\ee
up to order $1/N$.
To proceed further we consider the expression
\beq
\int_{-\infty}^{\infty}\left[\smat{\ln a_\pm\cr\ln\ol{a}_\pm }^\prime
  (\ln A_\pm\;,\ln\ol{A}_\pm)-\smat{\ln a_\pm\cr\ln\ol{a}_\pm }
  (\ln A_\pm\;,\ln\ol{A}_\pm)^\prime\right]dx 
\label{expr}
\eeq
which can be written exactly as
\beq
L(z)+L(1/z)={\pi^2\over 3} \label{L+L}
\eeq
in terms of the Rogers dilogarithmic function  
\beq
L(x)=\int^x_0\left({\ln(1+y)\over y}-{\ln y\over 1+y}\right]dy\label{L(x)}.
\eeq

On the other hand, making use of (\ref{a12}) in (\ref{expr}) and using 
$a_\pm(-\infty)=\overline{a}_\pm(-\infty)=0$
and $a_\pm(\infty)=1/\overline{a}_\pm(\infty)=e^{i(\omega_\pm-\phi)}$,
we arrive at the result
\beq
\mp 8\sqrt{3}\im\left(e^{\mp 2\rho\i\xi} \int_{-\infty}^{\infty}
      \ln A_\pm(x)e^{-x}\right) +{\pi(\omega_\pm-\phi)^2\over\pi-2\lam}
.\label{327}
\eeq  
Equating (\ref{327}) and (\ref{L+L}) gives the integral in (\ref{T1}). Thus
inserting this integral into the expression (\ref{T1}) 
we obtain 
\beq
\ln T(v)=-Nf_\infty(v)-{\pi\sin(2\i\rho v)\over 6N}(c-24\Delta)
\eeq
to leading order in $1/N$. This is our final result, from which
the central charge and conformal weights can be read-off
\cite{BCN:86,Affleck:86,cx} as 
\be
c&=&1-{3 \phi^2\over \pi(\pi-2\lam)}, \label{c12} \\
\Delta&=&\cases{\disp{(\omega-\phi\mp 2\ell\lam)^2
   -(\pi-4\lam)^2\over 8\pi(\pi-2\lam)} & dilute O($n$) model,\cr
\disp{(\omega-\phi)^2
   -(\pi-4\lam)^2\over 8\pi(\pi-2\lam)}  
 & dilute $A_L$ face model.\cr}\label{res12}
\ee

\section{Branches 3 and 4}
\setcounter{equation}{0}

On branches 3 and 4 the spectral and crossing parameters are
specialized in the regime
\beq
-\pi+3\lam<u<0\h\h {1\over 6}\pi\le\lam<{1\over 3}\pi\;.
\eeq
The following computation of the finite-size corrections to the 
transfer matrix eigenspectra for each of the models is valid for the 
larger interval $0<\lam<{1\over 3}\pi$.

We proceed in a similar manner as for branches 1 and 2 and
introduce a new parameter $v=iu$ and set $v_j=u_j$.
The function $p(v)$ is defined by
\beq
p(v)=e^{\i(\omega-\phi)} 
 {\Phi(v- \i \lam+\half\pi\i)q(v-\i\lam)q(v+2\i\lam)
  \over\Phi(v+\i \lam+\half\pi\i)q(v+\i\lam)q(v-2\i\lam)},
\eeq
with $\Phi$ and $q$ as given in (\ref{phiq}). 
In \cite{WBN:92,Warnaar} it has been checked 
that the Bethe ansatz roots are (almost) located on the lines 
$\im(v)=\pm\half\lam$ in the complex $v$-plane.
As a consequence we still have the 
symmetries of equation (\ref{qp}).

\subsection{Nonlinear integral equation} 

We proceed by defining functions that
are {\sc anz} in the strips around the real axis:
\be
A(v)=1+\alpha(v)/g(v) &\h& \alpha(v)=g(v)p(v-\i\lam)[1+p(v)] \no\\
B(v)=1+\beta(v)/g(v)  &\h& \beta(v)=g(v)\disp{p(v)p(v-\i\lam)\over 
                           1+p(v-\i\lam)} \no\\
C(v)=1+\gamma(v)/g(v) &\h& \gamma(v)=g(v)p(v-\i\lam) \\
                      &\h& \delta(v)=p(v). \no
\ee
The function $g(v)=\mbox{th}^N\rho(v+\i\lam-\half\pi\i)$, with 
$\rho=\pi/(2\pi-6\lam)$, is introduced to compensate the anticipated bulk 
behaviour of the functions $\alpha,\beta,\gamma$.

We define the Fourier transform of the functions $\alpha,\beta,\gamma$ as
in (\ref{Fou}). For $q$ we have
\be
q(k)={1\over 2\pi}\disp{\int_{-\infty+{\rm i} r}^{\infty+\i r}}
 \(\ln q(v)\)^{\prime\prime}e^{-{\rm i} kv}\;dv
 &\hs{0.5}&-\pi+\half\pi<r<-\half\pi\no \\
\(\ln q(v)\)^{\prime\prime}=\disp{\int_{-\infty}^{\infty}}q(k)
     e^{\i kv}\;dk&\hs{0.5} &-\pi+\half\pi<\im(v)<-\half\pi \\
q_1(k)={1\over 2\pi}\disp{\int_{-\infty+{\rm i} r}^{\infty+\i r}}
          \(\ln q(v)\)^{\prime\prime}e^{-{\rm i} kv}\;dv 
   & \hs{0.5}&-\half\pi<r<\half\pi\no \\
\(\ln q(v)\)^{\prime\prime}=\disp{\int_{-\infty}^{\infty}}
    q_1(k)e^{\i kv}\;dk&\hs{0.5} &-\half\pi<\im(v)<\half\pi.
\ee

To solve the functional relations we need the relations of the Fourier
transforms of $\alpha,\beta,\gamma$. 
First we can see that not all functions 
are independent and thus we have
\be
\beta(k)-\gamma(k)-\delta(k)+C(k)&=&0   \no \\
\alpha(k)-\ol{\alpha}(k)-\gamma(k)
  +\ol{\gamma}(k)-\delta(k)&=&0  \label{rel-1}  \\
A(k)-B(k)-C(k)&=&0.   \no
\ee
Applying the Fourier transform to the $\delta,\gamma$ gives 
\be
\gamma(k)&=&Nk\sinh\lam k/\sinh{\pi k/2}
 -\half Nke^{k\lam\over 2}/\cosh {k\over 2}(3\lam-\pi) \no \\
  &&\h +(e^{-\lam k+\pi k}+  
   e^{2\lam k}-e^{3\lam})q(k)-q_1(k) \label{gamma-q}\\
\delta(k)&=& Nk\sinh\lam k/\sinh{\pi k/2}-
  4e^{k\pi\over 2}\sinh{\lam k/2}
 \cosh {k\over 2} (3\lam-\pi)\; q(k). \label{delta-q}
\ee
Other relations follow by applying Cauchy's 
theorem to the auxiliary functions
\be
h_1(v)&=&p(v-\half\i\lam)[1+p(v+\half\i\lam)], \label{h2} \\
h_2(v)&=&{1+p(v-\half\i\lam)[1+p(v+\half\i\lam)] 
 \over p(v+\half\i\lam)}, \label{h3} \\
h_3(v)&=&[1+p(v-\half\i\lam)
 [1+p(v+\half\i\lam)]/q(v-\half\i\lam),\label{h4}
\ee
which all satisfy the non-trivial analyticity property
\beq
\disp{\int_{-\infty+\half{\rm i}\lam}^{\infty+\half{\rm i}\lam}}
\(\ln q(v)\)^{\prime\prime}e^{-{\rm i} kv}\;dv=
  \disp{\int_{-\infty-\half{\rm i}\lam}^{\infty-\half{\rm i}\lam}}
 \(\ln q(v)\)^{\prime\prime}e^{-{\rm i} kv}\;dv\;.
\eeq
It follows, respectively, that
\be
\alpha(k)&=&-e^{\lam k}\ol{\beta}(k) ,\label{alphabeta} \\
A(k)-\delta(k)&=&e^{\lam k}[\ol{A}(k)+\delta(k)],
\ee
and
\be
&&e^{-\half\lam k}\left(A(k)-e^{\lam k}q(k)\right) \no \\
&&\;\;=\;e^{\half\lam k}\left(\ol{B}(k)-\ol{\beta}(k)-q_1(k)+{\half Nk
  e^{-\half\lam k}\over\cosh 
(\talf\lam k-\half\pi k) }\right)\;.\label{rel-4}
\ee

Now solving (\ref{rel-1})--(\ref{rel-4}) and their complex conjugates 
in terms of the functions $A$ and $B$, we find
\be
\alpha(k)+\gamma(k)&=&F(k)A(k)+G(k)\ol{A}(k)+H(k)B(k)
 +\ol{H}(k)\ol{B}(k) \no\\
\beta(k)-\gamma(k)&=&\ol{H}(k)A(k)+\ol{H}(k)\ol{A}(k)+B(k)  \no\\
C(k)&=&A(k)-B(k) \label{q34}\\
&&\hs{-2.5}q(k)={Nke^{-\half\pi k}
  \cosh{\lam k\over 2}\over 2\sinh{\half\pi k}
  \cosh{k\over 2}(3\lam-\pi)}-{e^{-\half(\pi+\lam) k}A(k)
        -e^{-\half(\pi-\lam) k}\ol{A}(k)\over 4\sinh{\lam k}
  \cosh{k\over 2}(3\lam-\pi) } \no
\ee
with
\be
F(k)&=&{\sinh {k\over 2}(\pi-3\lam) -2 \sinh {k\over 2}(\pi-5\lam)\over
   2\sinh\lam k \cosh {k\over 2}(3\lam-\pi)} \no \\
G(k)&=& {3e^{-\half(\pi-5\lam) k}-2e^{-\half(\pi-7\lam) k}
    -2e^{-\half(\pi-3\lam) k}+e^{\half(\pi-\lam) k}  \over
   4\sinh\lam k \cosh{k\over 2}(3\lam-\pi)}   \\
H(k)&=& -{e^{-\half\lam k}\over 2  \cosh \half\lam k}.
\ee
Transforming back and integrating twice, we obtain 
a coupled set of nonlinear integral equations,
\be
\ln \alpha(v)+\ln \gamma(v)&=&F*\ln A+G*\ln\ol{A}
     +H*\ln B+\ol{H}*\ln \ol{B}+C,\no\\
\ln \beta(v)-\ln \gamma(v)&=&\ol{H}*\ln A
 +\ol{H}*\ln\ol{A}+\ln B, \label{nonli}
\ee
where we have introduced
\be
F(v)&=&{1\over 2\pi}\int_{-\infty}^\infty F(k) e^{\i k v}\;dk, \no \\
G(v)&=&{1\over 2\pi}\int_{-\infty}^\infty G(k) e^{\i k v}\;dk,  \\
H(v)&=&{1\over 2\pi}\int_{-\infty}^\infty H(k) e^{\i k v}\;dk. \no 
\ee
Taking the scaling limit as in (\ref{scaling12}), the nonlinear integral
equations become
\be
\smat{\ln a_{\pm}+\ln c_{\pm}\cr\ln b_{\pm}-\ln c_{\pm}\cr
      \ln \ol{a}_{\pm}+\ln \ol{c}_{\pm}\cr
      \ln \ol{b}_{\pm}-\ln \ol{c}_{\pm}}=\pm 4\i e^{-x}
 \smat{-e^{\pm \rho\i\lam}\cr 0\cr -e^{\mp\rho\i\lam}\cr 0}
 +K*\smat{\ln A_\pm\cr\ln B_\pm\cr\ln\ol{A}_\pm\cr\ln\ol{B}_\pm}
 +C_\pm\smat{1\cr0\cr -1\cr 0}
  \; ,\label{a34}
\ee
where the kernel $K$ again satisfies the very useful symmetry 
$K^T(x)=K(x)$.
The integration constant in  (\ref{nonli}) follows from the 
limit $x\to\infty$. We have 
\beq
C_\pm= {\pi\i(\omega_\pm-\phi)/( 2\lam)}.  
\eeq
Again the integration constants $C_\pm$ contain the essential information
to obtain the conformal spectra.

\subsection{Conformal spectra} 

For branches 3 and 4 the transfer matrix eigenvalues in  
(\ref{BAE}) are dominated by
\be
T(v+\i\lam-\half\pi\i)\sim e^{\i\phi} 
  \Phi(v+\i\lam+\half\pi\i)\Phi(v+\half\pi\i)
    {q(v-3\i\lam)\over q(v-\i\lam)}\;A(v).
\ee
Taking Fourier transforms and
using the solution (\ref{q34}) for $q(k)$, we obtain
\be
&&\ln T(v+\i\lam-\half\pi\i)\;=\;-N f_\infty(v)\no \\
 &&\h+\; {\rho\over \pi}\int_{-\infty}^{\infty}
 \left({\ln A(w)\over\cosh 2\rho(v-w+\half\i\lam)}
  - {\ln \ol{A}(w)\over\cosh 2\rho(v-w-\half\i\lam)} 
   \right)dw,
\ee
where the bulk free energy is given by
\be
\hs{-1} f_\infty(v)=2\!\int_{-\infty}^{\infty}\! dk {\sinh( k\i v)
   \sinh(3k\lam+\i vk-\pi k)
   \cosh(5k\lam-k\pi)\cosh(k\lam)\over k\cosh(3k\lam-\pi k)\sinh(k\pi)} . 
\ee
Taking the thermodynamic limit $N\to\infty$ and using the definition as in
(\ref{scaling12}) gives
\be
\ln T(v)&=&-N f_\infty(v)+{2\i\over N\pi}e^{2\rho v}
   \im\left(e^{\rho\i\lam} \int_{-\infty}^{\infty}
                        \ln A_+(x)e^{-x}\right) \no \\
  &&\h\h +{2\i\over N\pi}e^{-2\rho v}
 \im\left(e^{-\rho\i\lam} \int_{-\infty}^{\infty}\ln 
      {A}_-(x)e^{-x}\right)dx.
\ee
To calculate the integral, we consider the expression
\be
\int_{-\infty}^{\infty}\left[
 \smat{\ln a_{\pm}+\ln c_{\pm}\cr\ln b_{\pm}-\ln c_{\pm}\cr
      \ln \ol{a}_{\pm}+\ln \ol{c}_{\pm}\cr
      \ln \ol{b}_{\pm}-\ln \ol{c}_{\pm}}^{\prime}
 \smat{\ln A_\pm\cr\ln B_\pm\cr\ln\ol{A}_\pm\cr\ln\ol{B}_\pm}^T
 -\smat{\ln a_{\pm}+\ln c_{\pm}\cr\ln b_{\pm}-\ln c_{\pm}\cr
      \ln \ol{a}_{\pm}+\ln \ol{c}_{\pm}\cr
      \ln \ol{b}_{\pm}-\ln \ol{c}_{\pm}}
 \smat{\ln A_\pm\cr\ln
B_\pm\cr\ln\ol{A}_\pm\cr\ln\ol{B}_\pm}^{\prime\; T}\right]
 dx, \label{express}
\ee
which can be evaluated exactly using the Rogers dilogarithmic
function relation (\ref{L+L}). We thus arrive at
\be
&&L\(a_\pm(\infty)\)+L\(1/a_\pm(\infty)\)
 +L\(1/b_\pm(\infty)\) \no \\ 
&&\h +L\(b_\pm(\infty)\)
  +L\(c_\pm(\infty)\)+L\(1/c_\pm(\infty)\)=
  \pi^2-8k\pi^2, \label{hand-1}
\ee
where $k=0,1$ \cite{KWZ:93}. 
We have also used the asymptotics of the functions  
$a_\pm(\infty)=e^{i(\omega_\pm-\phi)}(e^{i(\omega_\pm-\phi)}+1)$, 
$b_\pm(\infty)={ e^{2i(\omega_\pm-\phi)}/(e^{i(\omega_\pm-\phi)}+1)}$, 
$c_\pm(\infty)=e^{i(\omega_\pm-\phi)}$ and $a_\pm(-\infty)=c_\pm(-\infty)
=c_\pm(-\infty)=0$.  

On the other hand, substituting  (\ref{a34})
into (\ref{express}) we arrive at 
\be
\pm 16\im\left(e^{\pm\rho\i\lam} \int_{-\infty}^{\infty}
                        \ln A_\pm(x)e^{-x}\right)
  +{\pi(\omega_\pm-\phi)^2\over\lam}.\label{hand-2}
\ee  
Combining the results (\ref{hand-1}) and (\ref{hand-2}) we are left with
\be
\im\left(e^{\pm\rho\i\lam} \int_{-\infty}^{\infty}\ln A_\pm(x)e^{-x}\right)
=\pm{\pi^2\over 24}\left({3\over 2}-{3(\omega_\pm-\phi)^2\over
2\pi\lam}- 12 k\right).
\ee
Inserting this in the expression $\ln T(v)$ we obtain 
the final result
\be
\ln T(v+\i\lam-\half\pi\i)=-Nf_\infty(v)+{\pi\sin(2\i\rho v)\over 6N}
  (c-24\Delta).
\ee
The central charges and conformal weights are given by
\be
c&=&{3\over 2}-{3(\pi-4\lam)^2\over 2\pi\lam} ,\label{c34} \\
\Delta&=&\cases{\disp{(\omega-\phi\mp 2\ell
   \lam)^2-(\pi-4\lam)^2\over 16\pi\lam}+\Delta_{\rm Ising}    
     & dilute O(n) model \cr
\disp{(\omega-\phi)^2-(\pi-4\lam)^2\over 16\pi\lam}
          +\Delta_{\rm Ising}  & dilute $A_L$ face model\cr} \label{res34}
\ee
with $\Delta_{\rm Ising}=0,{1\over 2}$.

\section{Summary and discussion}

We have calculated the finite-size corrections to the transfer matrix 
eigen-spectra of the intimately related dilute O($n$),  
dilute $A_L$ and Izergin-Korepin models at criticality
via the nonlinear integral equation approach.
The resulting conformal weights defining the critical
exponents are seen to follow from appropriate branches
of the log functions appearing in the Bethe equations.
For the dilute O($n$) model the integration constants appearing
in the key nonlinear integral equations (\ref{a12}) and (\ref{a34}) 
differ in the distinct limits $v\to\pm\infty$,
leading to two constants, $C_{\pm}$, in the scaling limit.
However, they satisfy $|C_+|=|C_-|$, allowing the calculation
to go through in a similar manner as for the ABF model \cite{Zhou:95}.

\subsection{Dilute O($n$) model}

Consider first the dilute O($n$) model in branches 1 and 2 in
the $u$ positive regime.
Our final results are the central charge (\ref{c12}) and 
the conformal weights (\ref{res12}). 
These are in agreement with the previous results outlined in
Sections 2.1 and 2.2 with the O($n$) $\phi$ value (\ref{on}). 
In particular, the magnetic dimensions (\ref{mag12}) follow 
with the parameters $\ell\ne 0$ and $r=0$ in (\ref{casespm}).
Similarly, the thermal dimensions (\ref{therm12}) follow with 
$\ell= 0$ and $r=2j$. 
As expected, in this
regime the conformal dimensions are seen to be in agreement
with the results obtained in the honeycomb limit. 

On the other hand, in branches 3 and 4 of the $u$ negative regime
our final results are (\ref{c34}) and (\ref{res34}). 
Here the conformal dimensions are new. The conjectured
magnetic dimensions (\ref{mag34}) \cite{WBN:92} are associated with 
$\ell\ne 0$ and $r=\ell$.
The similarly conjectural thermal dimension 
$X_1^\epsilon =2\Delta_{\rm Ising}=1$  
follows from the choice $\ell= 0$ and $r=0$.
More generally we see that this dimension belongs to the
thermal set
\be
X_{j+1}^\epsilon={j^2\pi-j(\pi-4\lambda)\over 2\pi\lambda}
    +\Delta_{\rm Ising},
\ee 
which is given by setting $\ell=0$ and $r=2j$ in (\ref{res34}).

\subsection{Dilute $A_L$ face model}

Recall that the Bethe equations of the dilute $A_L$ face model 
follow from the choice of crossing parameter given in (\ref{branches})
with seam $\phi$ as given in (2.5). In this case the branches 1 and 2
results (\ref{c12}) and (\ref{res12}) give the central charge (2.11)
and conformal weights (2.15) of the unitary minimal series.

In a similar manner, the branches 3 and 4 results are indicative
of the product of the unitary minimal series with the Ising model. 
The central charge is given by (2.11) and the conformal weights
again by given by (2.15), with however, the additional 
$\Delta_{\rm Ising}$ component. Here $\Delta_{\rm Ising} = 
0,\frac{1}{2},\frac{1}{16}$ is expected, in accordance with the
results of \cite{WPSN:94}. However, $\frac{1}{16}$ does not appear 
in our results. Our method needs further refinement to reveal this
conformal dimension.  

For the more general crossing parameter 
$$ \lam={\pi\over 4}(1+{k\over L+1}) $$
with $|k|<\lfloor (L+1)/3\rfloor$, the integer part of
the fraction $(L+1)/3$, our results (\ref{c12}) and (\ref{c34}),
and (\ref{res12}) and (\ref{res34}) imply 
\be
c&=&1-{6k^2\over (L+1)(L+1-k)}  \h\mbox{branches 1 and 2}\\
c&=&{3\over 2}-{6k^2\over (L+1)(L+1-k)} \h\mbox{branches 3 and 4}
\ee
for the central charges and 
\be
\Delta&=&{[(L+1)t-(L+1-k)s]^2-k^2\over 4(L+1)(L+1-k)} 
               \hs{2.0}\mbox{branches 1 and 2}\\
\Delta&=&{[(L+1)t-(L+1-k)s]^2-k^2\over 4(L+1)(L+1-k)}+\Delta_{\rm Ising} 
             \h\mbox{branches 3 and 4} 
\ee
for the conformal weights, where 
$s=1,2,\cdots,L$ and $t=1,2,\cdots,L-k$. These results indicate the
non-unitary minimal models for the dilute $A_L$ models in braches 1 and 2
and the non-unitary minimal models plus an Ising model 
in branches 3 and 4.  For the 
case  $k=1$ our results confirm the conformal weights presented 
in \cite{WPSN:94}. For $k>1$ our results show that  
the dilute $A_L$ models in branches 1 and 2 can be classified
by the same universality
classes as for the ABF models  \cite{FoBa:85,Zhou:95}.

\subsection{Izergin-Korepin model}

Our results for the O($n$) model reduce to those of the
Izergin-Korepin model when the seam $\phi=0$. 
In this way the central charge and conformal weights are given by 
\beq
c = 1, \quad X_{\ell,r} =
  \disp{\ell^2(\pi- 2\lam)\over 4\pi}+
 \disp{r^2\pi\over \pi-2\lam} \label{iz12}
\eeq
for $0<u<3\lam$ with  $0 <\lam < \pi/3$. On the other hand,
\beq
c = {3\over 2}, \quad X_{\ell,m} =
 \disp{\lam\ell^2\over2\pi}
  +{m^2\over 2\pi\lam} +2 \Delta_{\rm Ising} \label{iz34}
\eeq
for $-\pi+3\lam<u<0$ with $0< \lam < \pi/3$. 
The result (\ref{iz12}) is in agreement with the previous results in
the honeycomb limit, while the result (\ref{iz34}), reflecting also the
additional Ising content, is new. 

\vskip 0.5cm

It is a pleasure to thank Ole Warnaar for helpful discussions 
over the years. 
This work has been supported by the Australian Research Council.
YKZ also thanks the Natural Science Foundation of China
for partial support.


\end{document}